\documentclass[preprint,12pt,numbers,authoryear]{elsarticle}

\usepackage[utf8]{inputenc}
\usepackage[T1]{fontenc}
\usepackage[english]{babel}
\usepackage{amssymb}
\usepackage{float}
\usepackage{fancyhdr}
\usepackage{natbib}
\usepackage{tablefootnote}
\usepackage{graphicx}        
\usepackage{color}           

\usepackage{breakurl}        

\usepackage[squaren]{SIunits}
\newcommand{\micron}{\ensuremath{\text{{\micro}m}}}
\newcommand{\moire}{moir\'e}

\newcommand{\Def}[1]{\emph{#1}}  

\usepackage{amsmath} 

\fancypagestyle{pprintTitle}{%
\lhead{\footnotesize{https://doi.org/10.1016/j.asr.2019.08.040 \\ \textcopyright 2019 This manuscript version is made available under the CC-BY-NC-ND 4.0}}
\lfoot{}\cfoot{}\rfoot{}

}

\begin{document}

\begin{frontmatter}

\title{The Micro Solar Flare Apparutus (MiSolFA) Instrument Concept}

\author[aff1,aff2]{Erica Lastufka}
\ead{erica.lastufka@fhnw.ch}
\author[aff1,aff4]{Diego Casadei}
\ead{diego.casadei@cosylab.com}
\author[aff1,aff3]{Gordon Hurford}
\ead{ghurford@ssl.berkeley.edu}
\author[aff1,aff2]{Matej Kuhar}
\ead{matej.kuhar@fhnw.ch}
\author[aff1,aff5]{Gabriele Torre}
\ead{gabriele.torre@kellify.com}
\author[aff1,aff6]{S\"am Krucker}
\ead{krucker@berkeley.edu}

\address[aff1]{University of Applied Sciences Northwest Switzerland, Bahnhofstrasse 6, CH-5210 Windisch, Switzerland}
\address[aff2]{ETH-Z\"urich, R\"amistrasse 101, CH-8092 Z\"urich, Switzerland}
\address[aff3]{Space Sciences Laboratory, University of California, Berkeley, CA 94720-7450, U.S.A.}
\address[aff4]{Cosylab Switzerland GmbH, CH-5234 Villigen, Switzerland}
\address[aff5]{Kellify s.p.a, IT-16121 Genova, Italy}
\address[aff6]{Space Science Laboratory, UC Berkeley}

\begin{abstract}
The Micro Solar-Flare Apparatus (MiSolFA) is a compact X-ray imaging spectrometer designed for a small 6U micro-satellite. 
As a relatively inexpensive yet capable Earth-orbiting instrument, MiSolFA is designed to image the high-energy regions of solar flares from a different perspective than that of Solar Orbiter's STIX, operating from a highly elliptical heliocentric orbit. 
Two instruments working together in this way would provide a 3-dimensional view of X-ray emitting regions and can bypass the dynamic range limitation preventing simultaneous coronal and chromospheric imaging.
Stereoscopic X-ray observations would also contain valuable information about the anisotropy of the flare-accelerated electron distribution. 
To perform these types of observations, MiSolFA must be capable of imaging sources with energies between 10 and 100 keV, with 10 arcsec angular resolution.

MiSolFA's Imager will be the most compact X-ray imaging spectrometer in space. Scaling down the volume by a factor of ten from previous instrument designs requires special considerations. Here we present the design principles of the MiSolFA X-ray optics, discuss the necessary compromises, and evaluate the performance of the Engineering Model.

\end{abstract}


\end{frontmatter}
\thispagestyle{fancy}

\section{Introduction}\label{sec-intro} 

\begin{figure}[t]
  \centering
  \includegraphics[scale=0.45]{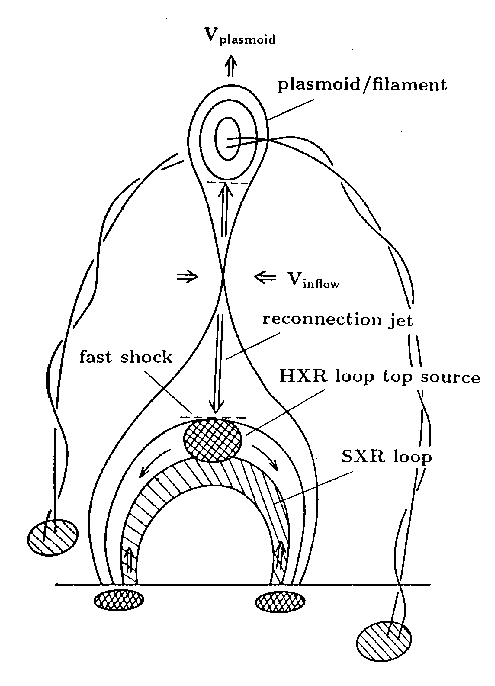}
  \caption{Illustration of the standard model of solar flares and the acceleration of electrons due to magnetic reconnection from \citet{shibataEvidenceMagneticReconnection1998}. Hard X-ray sources can be observed at the loop top and footpoints, while sometimes the flare loops emit soft X-rays.}
  \label{fig-cme}
\end{figure}

Solar flares are powerful events capbale of releasing up to $10^{32}$ ergs of energy in a few minutes \citep{2008LRSP....5....1B,2011SSRv..159..107H}.
A large fraction of this energy, previously stored in coronal magnetic fields and suddenly released by magnetic reconnection, goes into the acceleration of particles. 
Electrons become observable as X-rays, emitted when they collide with the ambient plasma to produce bremsstrahlung radiation. The standard flare model (Figure \ref{fig-cme}) describes two main sites where this occurs. One is in the corona, where accelerated particles collide with plasma-filled flare loops. The other is lower in the chromoshpere, where electrons travelling down closed magnetic field lines collide with the very dense chromospheric material at the loop "footpoints".

The observed X-ray spectrum depends not only on the energy distribution of the electrons, but also on their pitch angle distribution \citep[e.g.][]{2011SSRv..159..301K,2004ApJ...613.1233M}. 
A possible implication of the standard flare model is that, due to the downward flux of energetic electrons toward the chromosphere \cite[e.g.][]{1971SoPh...18..489B}, X-rays are emitted along a preferential direction - contrary to all measurements made so far. 
Assuming an injected or radiated electron population with isotropic or anisotropic electron pitch-angle distribution can profoundly impact inferred solar plasma parameters \citep{2014ApJ...787...86J}. 

\begin{figure}[t]
  \centering
  \includegraphics[width=\textwidth]{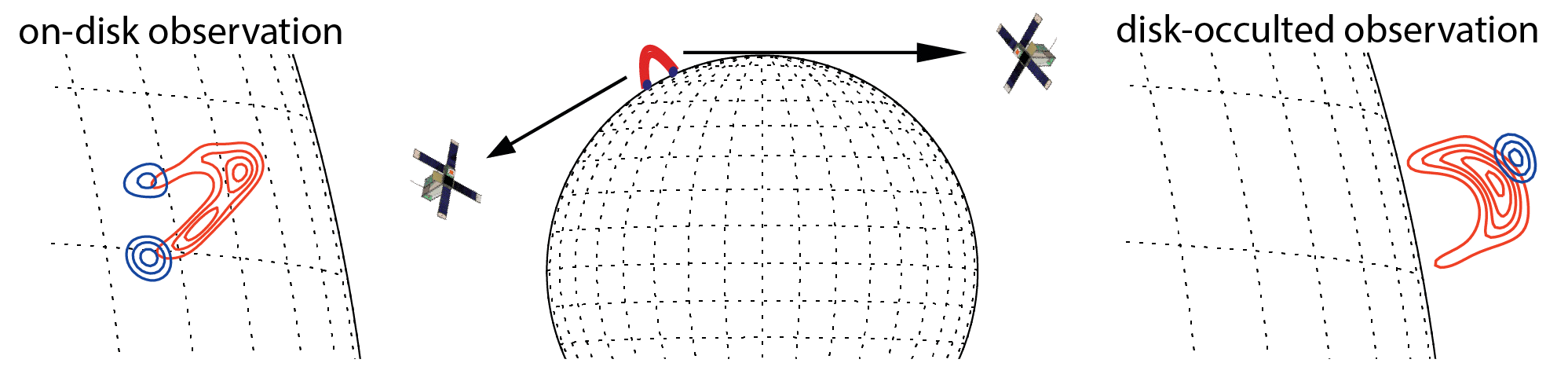}
  \caption{Two satellites use occultation by the solar disk to observe different X-ray emitting regions of the same flare. An on-disk observation shows the hard X-ray footpoints (blue) and possibly the soft X-ray flare loop (red). When the bright footpoints are occulted, the soft X-ray loop can be seen without interference, and any looptop hard X-ray source (blue) becomes visible. }
  \label{fig-occ}
\end{figure}

The hard X-ray (HXR) spectrum above few tens of keV is a vital probe of the non-thermal physical processes occurring on the Sun
\citep[e.g.][]{2011SSRv..159..107H,2011SSRv..159..301K,2003ApJ...595L.115B}. However, current instrumentation (mostly using indirect imaging) finds it difficult to observe both the coronal and chromospheric X-ray sites simultaneously. A single indirect imager suffers from a limited dynamic range; from the Reuven Ramaty High Energy Solar Spectroscopic Imager \citep[RHESSI;][]{2002SoPh..210....3L}, we know that seeing two or more sources in the same field-of-view is usually only possible if their intensity differs by less than a factor of 5--10. 
 Because chromospheric sources are usually at least two orders-of-magnitude brighter than sources higher in the corona, coronal emission can only be studied if the bright footpoints are hidden behind the solar limb (see Figure \ref{fig-occ}, right side).
 
 A pair of satellites could take advantage of such partially-occulted observations; one could image the whole flare, dominated by footpoint emission, and the other the coronal emission (Figure \ref{fig-occ}). 
Even without utilizing occultation, a satellite pair with cross-calibrated detectors and different points-of-view could make unambiguous measurements of solar flare electron anisotropy 
\citep[e.g.][]{1981Ap&SS..75..163K,1986ASSL..123...73H,1988ApJ...326.1017K,1990ApJ...359..524M,1998ApJ...500.1003K, casadeiMeasuringXrayAnisotropy2017a}. Here, good spectral resolution and detector compatibility are more important than imaging capability. 
With the Spectrometer Telescope for Imaging X-rays  \citep[STIX;][]{benzSpectrometerTelescopeImaging2012} due to launch into a highly elliptical orbit aboard the Solar Orbiter \citep{2013SoPh..285...25M}, the time is ripe for a complementary Earth-orbiting X-ray instrument.
An imaging spectrometer with cross-calibrated detectors is ideal; from the imaging we could simultaneously study flare looptop and footpoint emission; with the detectors we could systematically quantify X-ray directivity for the first time.

In this paper, we present the Micro Solar Flare Apparatus (MiSolFA) as the best available instrument for these two tasks. Designed to fit into a 6U cubesat, MiSolFA can be launched into low Earth orbit, providing the needed stereo viewpoint to STIX. 
From Table \ref{tab-missions}, we see how the MiSolFA small satellite concept makes great strides in reducing mass and volume while maintaining high performance capability compared to previous instrumentation; it also maintains these advantages over upcoming missions. 

   \begin{table}[h]
  \centering
  \resizebox{\textwidth}{!}{%
  \begin{tabular}{lccccl}
    Mission/Year  & Volume & Instrument Mass & Effective Area & Resolution & Energy range \\
      & (m$^3$)& (kg) & (cm$^2$)& (arcsec)&(keV) \\
     \hline
     Yohkoh HXT (1991)\tablefootnote{\citet{kosugiHardXrayTelescope1991}} &1.4x0.4x0.4 &40& 70&5&15-100\\ 
     RHESSI (1998) &2.2x1.2x1.2&293&50&2&3-17000\\
     STIX (2020) &0.8x0.3x0.3&7&7&7&4-150\\
     ASO-S HXI (2022)&1.6x0.7x0.7&164&100&6&30-200\\
     MiSolFA&0.3x0.2x0.1&2&3&10&10-100\\
     FOXSI (2025) &15x1x1&117&50&8&4-50\\
      \hline
  \end{tabular}}
  \caption{Comparison of MiSolFA with previous, upcoming and proposed Sun-observing hard X-ray imaging instruments. The mass given for RHESSI is of the entire spacecraft.}
  \label{tab-missions}
 \end{table}

With the retirement of RHESSI in 2018, there is currently no solar-dedicated hard X-ray imager in space. STIX will launch in 2020 and start providing science data at the end of a 2-3 year cruise phase. HXI is a current candidate for joint observations with STIX, and focusing optics instruments are an even more exciting option in the future.
The Advanced Space-based Solar Observatory with the Hard X-ray Imager (ASO-S HXI; \citet{ganASOSAdvancedSpacebased2015}) aboard is also scheduled for launch in 2022; however, it is only sensitive to large flares emitting above 30 keV. 
High-quality stereoscopic X-ray imaging together with STIX could still be possible, but detector cross-calibration would be challenging, especially at lower energies. The Focusing Optics X-ray Solar Imager (FOXSI: \citet{christeFocusingOpticsXray2016}), could fly on a NASA Medium-Class Explorer payload in low Earth orbit with earliest launch in 2025, has optics which allow a single instrument to escape the dynamic range limitation of indirect HXR imagers. 
As FOXSI uses CdTe detectors with similar spectral resolution to those of STIX, cross-calibration accuracy of the two instruments is expected to be good enough for directivity measurements. 

While MiSolFA is not the only instrument suitable for its purpose, its small size brings a major advantage: it is the fraction of the cost of the others. 
Here, we briefly summarize the satellite concept (section \ref{sec-sat}), before concentrating on the Imager.
Principles of the indirect imaging method and specific choices made to optimize it for a small satellite are presented in section \ref{sec-design}. 
The design and first performance of the engineering model (EM) of the MiSolFA Imager are reported in sections \ref{sec-EM} and \ref{sec-perf}, illustrating that the existing technology is indeed capable of producing a small yet powerful hard X-ray imager.

\section{The MiSolFA Satellite concept}\label{sec-sat}

The size of an indirect X-ray imager is driven by the separation distance between the two gratings placed in front of photon detectors. This, in turn, is driven by the smallest angular scale sampled by the instrument which is ultimately determined by the imaging requirements and the technological capability to produce fine-pitch diffraction gratings. A smaller instrument needs gratings with pitches down to tens of microns which are also thick enough to absorb high-energy X-rays.

We will show later that current technologies can produce thick gratings as fine as 15 \micron; for MiSolFA that means a reduction in grating separation from 1 m (RHESSI) to 15 cm, allowing the instrument to fit lengthwise in three one-liter cubesat units. Additional electronics take up another 3U. A possible block scheme of a 6U satellite equipped with the MiSolFA Imager is shown in Figure 3.
 Suitable existing technologies for the spacecraft bus include NASA's Dellingr platform. 

A satellite hosting the MiSolFA Imager requires a few components besides the instrument itself. A thermal shield, fixed to avoid the attitude complications associated with moving parts, should be placed in front of the Imager. 
An aspect system capable of finding the Sun position inside the field of view within few arcsec is also needed. This should be built into the Imager unit itself, to ensure the alignment of the instrument and the aspect system. 

Photon detectors placed directly behind the Imager are responsible for recording the counts which will eventually be converted to science data. For MiSolFA, the first candidate is the Caliste-SO CdTe detectors used by STIX \citep{meurisCalisteSOXraySpectrometer2014}. These have a surface area of 1~cm$^2$ and are composed of pixels with areas from 1 to 8 mm$^2$. Each half of the detector has four large pixels with a small pixel inset in one corner, for a total of eight large pixels and eight small pixels. 
With high-resolution photon detectors, requirements on the grating quality can be eased, which would benefit an instrument the scale of MiSolFA. 
Next-generation units based on hybridization techniques are under development which would reduce the pixel size to 250 \micron{}. 

The mass of the Imager itself, based on the prototype Engineering Model (EM), is around 1 kg. The total mass of a 6U micro-satellite carrying the Imager is not expected to exceed 5 kg. A small mass and volume makes such a satellite an ideal candidate for a launch as a secondary payload. To maximize scientific productivity, MiSolFA should be launched close to solar maximum in 2024, preferably in a polar orbit. 

\begin{figure}\label{fig-block-scheme}
\centering
\includegraphics[width=0.5\textwidth]{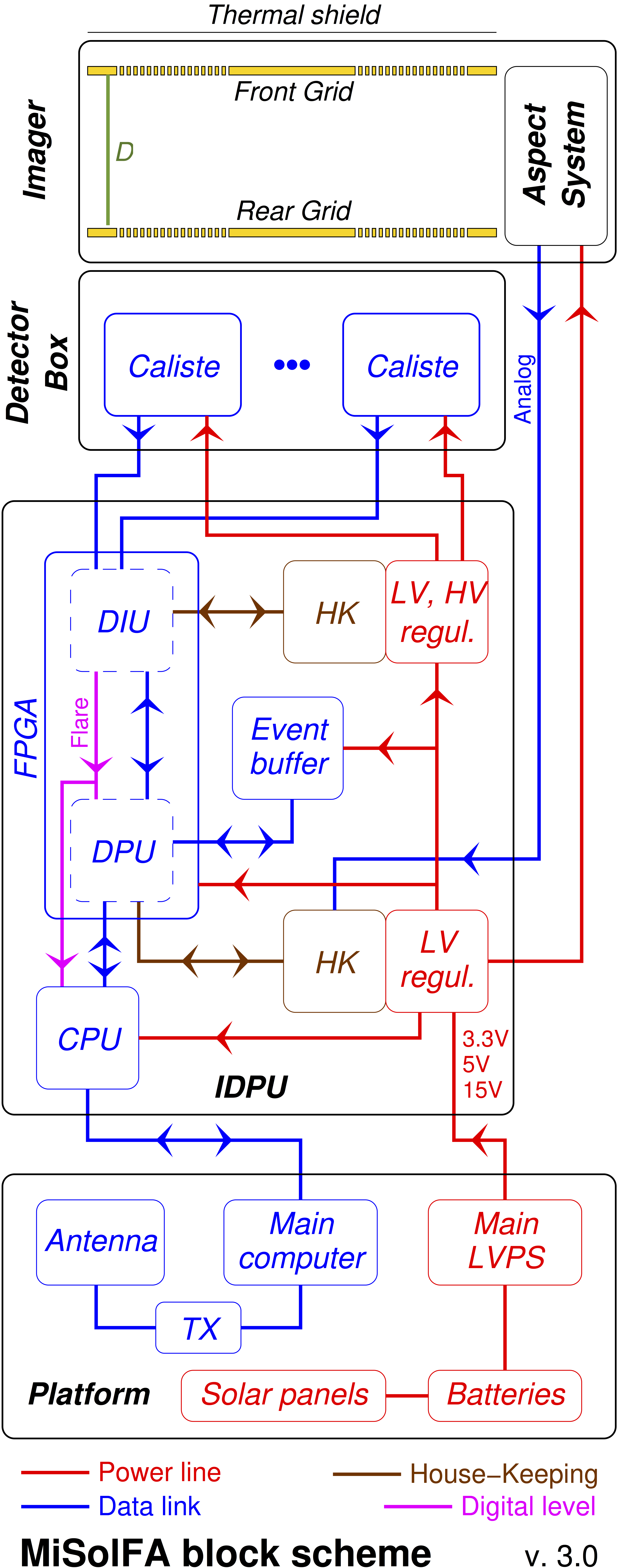}
\caption{Possible logistical block scheme of the MiSolFA satellite, where Caliste units are used as photon detectors and the Detector Interface Unit is separate from the Data Processing Unit. The Imager consists of a front and rear grid separated by a distance $D$, placed in front of photon detectors.}
\end{figure}

\section{Design of the MiSolFA Imager}\label{sec-design}

To achieve MiSolFA's science goals, we require an imaging system capable of separating two flare footpoints. In the case of occulted observations, the imaging should  enable us to determine the size of a coronal source. Chromospheric HXR footpoints are very compact, as they are often aligned along narrow flare ribbons visible in the EUV, and have generally been imaged at sizes below 10" \citep{dennisHARDXRAYFLARE2009a}. Thus our imaging goal is not to resolve the footpoints, but to separate them.
\citet{2008SoPh..250...53S} performed a statistical analysis of the footpoint separation of flares observed by RHESSI, 
and found that most events had a separation between 20'' and 30'', although a sizable fraction of events reached 40"--50''.
Coronal sources are larger, with even small flares exhibiting thermal loops with lengths around 30" \citep{hannahRHESSIMicroflareStatisticsII2008}. Therefore, the MiSolFA science requirements can be met by measuring Fourier components over a relatively narrow range of angular scales (10" -- 60") and using indirect imaging.  

The MiSolFA Imager's field of view includes the entire solar disk. By sampling different points in the Fourier plane, it produces visibilities which can be reconstructed into images of the X-ray source distribution. 
The choice of sampled angular scales determines the maximum detectable source size and the minimum separation required between two sources to distinguish them from one another.

One method of indirect imaging, also used by the STIX imaging system, exploits the moiré effect\footnote{RHESSI uses time modulation, as opposed to the spatial modulation of the \moire{} effect.}  \citep{princeGammarayHardXray1988}. 
When two gratings of slightly different pitch and orientation are placed one in front of the other,a periodic \moire{} pattern is formed on an instrument's photon detectors when X-ray flux is partially absorbed by the grating slats. The amplitude of this pattern is sensitive to the source size, with a characteristic angular scale fixed by half of the ratio between the average grating period and the separation between the front and rear grids.
Extended sources which are larger than the subcollimator angular scale make the moiré modulation disappear. Sources much smaller than the angular scale behave as point-like sources, producing the maximum modulation.
The \moire{} pattern phase changes with the source location or structure. 

We adopt the usual terminology for the absorption gratings. A \Def{grid} is the physical piece containing multiple gratings. Each individual grating occupies a square area called a \Def{window} on each grid.
A \Def{subcollimator} is the set consisting of a pair of front and rear windows, with a single corresponding photon detector placed directly behind the rear window. 
To accurately reconstruct a solar X-ray source, the subcollimators must be designed to sample a Fourier sequence of angular frequencies along multiple directions. 


\subsection{Grid Design}
 
In contrast to its predecessors RHESSI and STIX, the MiSolFA Imager prioritizes image quality over angular range. Operating from Low Earth Orbit means that angular sizes will be more constant than with STIX, and RHESSI statistical studies such as \citet{2008SoPh..250...53S} teach us what separation distances and energy distributions to expect from most solar flares.

\begin{figure}[t!]
   \begin{minipage}{0.5\linewidth}
     \centering
     \includegraphics[width=0.95\linewidth]{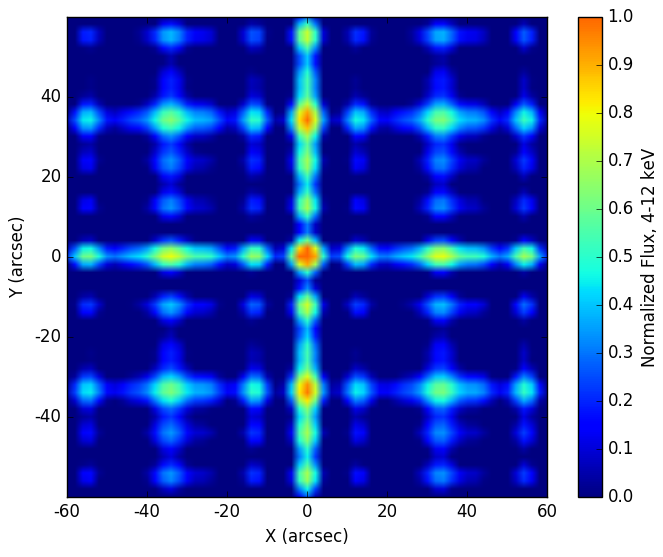}
   \end{minipage}%
   \begin{minipage}{0.5\linewidth}
     \centering
     \includegraphics[width=0.95\linewidth]{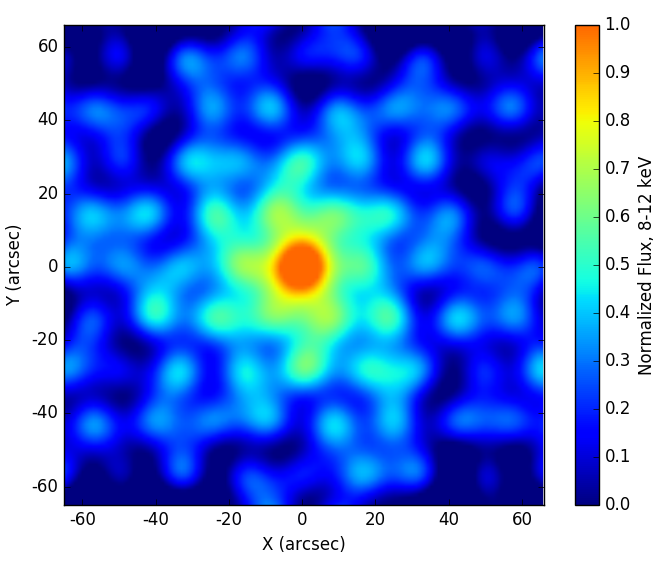}
   \end{minipage}
   \caption{Backprojection reconstruction of a point-source for
     MiSolFA (left) and STIX (right).}
   \label{fig-backproj}
 \end{figure}

The largest angular scale was chosen to be $a$=60", which should be sufficient to separate the footpoints from a looptop source in most flares. With a cubesat unit dimensions limiting grid width and height to 10 cm each, we can fit twelve windows with an area of 1 cm$^2$ on each grid, leaving 4 cm$^2$ for the aspect system. Each subcollimator provides one Fourier component of the X-ray image. The Fourier sequence of angular scales chosen for MiSolFA is therefore $a/n$, where $a$=60" and $n=1,2,\ldots6$. This samples angular scales of 60'', 30'', 20'', 15'', 12'', and 10''. Six subcollimators will provide 1-D information about the source from one direction. To build a 2-D picture, a second direction must be sampled by the remaining six subcollimators; by choosing subcollimators with grating orientations orthogonal to the orientations of the first six, we can sample the orthogonal direction. 
Because the largest angular scale is 60'', the resulting backprojection image (Figure~\ref{fig-backproj}, left plot) will repeat itself every 120'' along both directions. This ambiguity can however be resolved with the aid of other Sun-observing instruments, such as SDO/AIA. 

\begin{figure}[t]
   \centering
   \includegraphics[width=\textwidth]{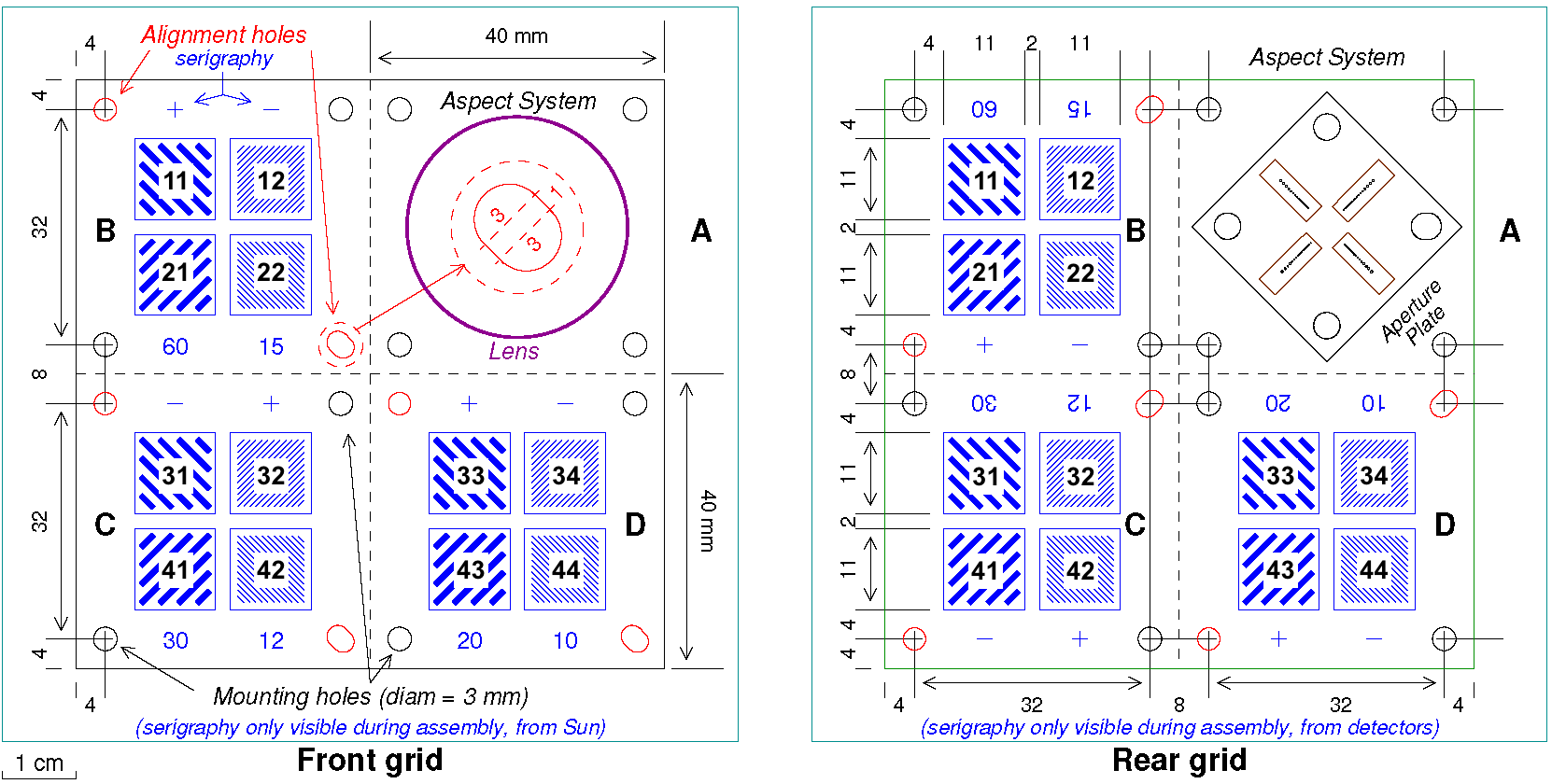}
   \caption{Design of the MiSolFA front and rear grids, with details specific to the Engineering Model (section \ref{sec-EM}). Dashed lines indicate the possible internal segmentation.}
   \label{fig-both-grids}
 \end{figure}

 As shown in Figure~\ref{fig-both-grids}, both front and rear grids consist of four \Def{segments}, all but one containing four windows each. The other segment contains the aspect system for determining absolute position. The arrangement of windows inside each segment, also illustrated in Figure \ref{fig-both-grids}, is optimized for mechanical integrity and to avoid increased background resulting from cross-talk. Having the grating orientations pointing to the edges of the segments in a radial fashion is more vibration-resistant than other options. Non-concurrent gratings with large differences in period placed opposite each other increases the probability that a photon passing through the coarse front grating in one window (for example, Window 11) at a wide enough angle to hit another window (for example, Window 12) will be absorbed by the finer rear slats, and vice versa. In any case, cross-talk effects are likely to be only on the order of a percent and can be corrected for in post-processing if the source location is well known.
 
 In order to simplify the production and reduce the cost of the imager, each segment with $2\times2$ windows is designed such that the segment located in the rear grid is obtained by flipping the corresponding front segment about the horizontal axis.


\subsection{Moir\'e period and orientation}\label{subsec-period}

 Each subcollimator has a front and rear grating which are plane-parallel to each other and separated by a fixed distance $D$ (see Figure 3). Each grating is a set of parallel, equally spaced absorbing slats separated by slits with a given period and orientation.  
 
\begin{figure}[t]
 \includegraphics[width=\textwidth]{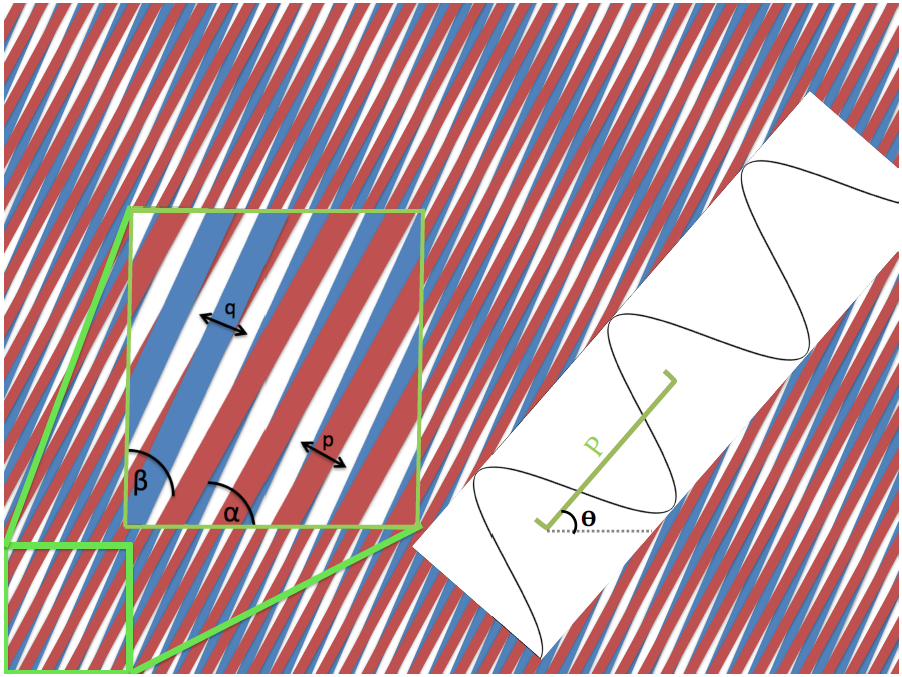}
  \caption{A grating of period $p$ and orientation $\alpha$ (red) is placed in front of a grating of period $q$ and orientation $\beta$ (blue, two slats of which are shown on top in the magnified area for clarity). The resulting low-frequency modulation, approximated by a cosine shown in the white inlay, is the \moire{} pattern which will appear on the detectors with period $P$ and orientation $\Theta$.}
  \label{fig-subcoll-concept}
 \end{figure}

For a given grating, the \Def{duty cycle} is defined as the ratio of the slit width to the grating period. The MiSolFA concept utilizes a duty cycle of 50\%, such that in the geometric approximation, the transmission of a single grating at any given energy is also 50\%.\footnote{The transmission is actually energy-dependent, and there may be diffraction effects.  Both aspects are ignored in this section, as they are on the order of a few percent.}
The transmission profile of a such a grating is a square wave alternating zero and full transmission at regular steps.
 The transmission envelope after the second regular grid is then a
 triangular distribution, enclosing a very rapid variation (with the same period, with good approximation). To the lowest order, we approximate the triangular wave with a cosine function. This is illustrated in Figure \ref{fig-subcoll-concept}, where the front grating has a period $p$ and orientation $\alpha$, and the rear grating has a period $q$ and orientation $\beta$. The eye is drawn to the large stripes of the \moire{} low-frequency modulation pattern, which has a period $P$ and orientation $\Theta$. A high-frequency modulation, produced by the crossings of the individual slats of the front and rear gratings, is also present but averages to zero for detectors of sufficient spatial resolution.
 
We require that MiSolFA's \moire{} pattern have a given period $P$, which runs vertically along each subcollimator. Specifically,
 \begin{enumerate}
 \item To achieve the desired angular scales $a$ with each subcollimator, we must choose $D$, $p$ and $q$ such that $a = S/(2D)$, with the geometric mean $S$ defined as $S^2 = qp$. Recall that we chose $a$ = 60", 30", 20", 15", 12" and 10". 
 \item The moiré period $P$ is close to the detector size. For MiSolFA, using Caliste-SO, this is 8.8 mm.
\item A \moire{} pattern running parallel or anti-parallel to the vertical axis has the property that the projections of slits in the front and rear gratings cross along lines parallel to the horizontal axis. These crossing points have uniform spacing $d$. 
 \end{enumerate}

 These requirements result in physical constraints on the front and rear grating periods $p$ and $q$ and orientations $\alpha$ and $\beta$, as well as the separation distance $D$ between the two.
To produce a \moire{} pattern that meets all these requirements, we solve for the particular periods and orientations of both gratings following the usual method \citep[e.g.][]{princeGammarayHardXray1988} and find as a first-order approximation:

\begin{equation}\label{alpha-beta-d-P-approx}
    \alpha, \beta = \frac{\pi}{4} \pm \frac{d}{4P}
\end{equation}
 
and

\begin{equation}\label{slit-periods-approx}
  p,q=S\left(1 \pm \frac{d}{4P}\right)
\end{equation}
 
 where
 
\begin{equation}\label{d-S-P-approx}
     d = \sqrt{2} S \left[ 1 - \left( \frac{S}{2P} \right)^4 \right]^{-1/4}
\end{equation}

From these equations, we can calculate the parameters defining MiSolFA's grids for a chosen separation distance $D$.

Additional requirements are important when designing the grids.
 If the pointing is only stable to within half a degree, a possible optimization would be to increase the duty cycle to slightly more than 50\%. 
  The high walls of the grating slats absorb wider-angle photons, such as those coming from flares near the solar limb, effectively reducing the slit width.
 Enlarging the slits, and therefore increasing the duty cycle, could help reduce the consequent loss in transmission.
 

 \subsection{Moir\'e phase}\label{subsec-phase}

\begin{figure}[t]
   \centering
   \includegraphics[scale=0.42]{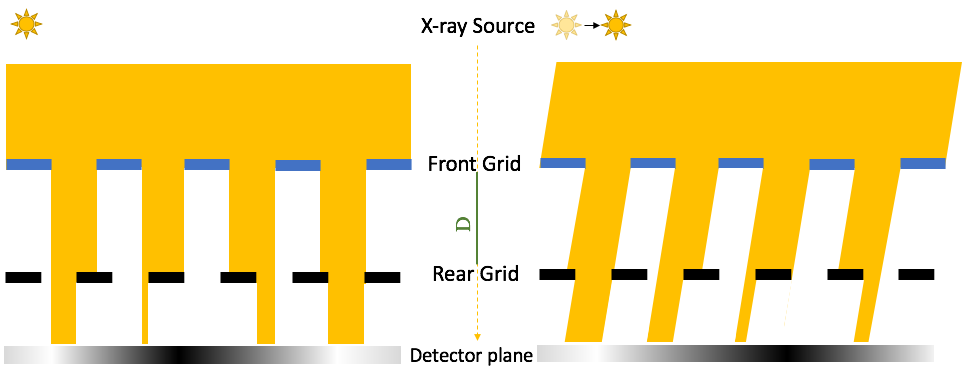}
   \caption{Moiré pattern shift corresponding to a source
     displaced to the right. The coarser front grid is represented by the blue dashes, and the finer rear grid by the black dashes, separated by a distance $D$. A source at the top of the figure at infinite distance away produces the black-and-white  intensity pattern shown on the left. This \moire{} pattern is shown as continuous instead of discrete to better reconcile the cartoon shown here with the observed patterns shown later. When the source is displaced, the intensity pattern shifts to the right.}
   \label{fig-moire-phase}
 \end{figure}

The \moire{} phase encodes information about the location of the X-ray source. 
Consider the diagram in Figure \ref{fig-moire-phase}, where an X-ray source is an infinite distance away from a pair of gratings, in a direction orthogonal to the plane containing the gratings. 
The coarse front and fine rear gratings are represented by the blue and black lines. 
At the source's original position, it produces the \moire{} pattern shown at the bottom of the top-left panel.
Displacing the source along the direction orthogonal to the grating orientation results in the \moire{} pattern shifting to the right, illustrated in the right panel (note the change in position of the dark \moire{} minimum).
If we exchange the front and rear grids, which is effectively done for half the MiSolFA subcollimators since the rear grid is the reflection of the front, we achieve the opposite effect; the \moire{} pattern moves to the left. 

 \section{The Engineering Model grids}\label{sec-EM}

Using the equations from \ref{subsec-period}, we optimized the instrument design to take into account physical size constraints, imaging requirements for science, and technological progress in fine-grating production. With the finest grating 15 \micron{} in pitch sampling 10", the instrument would have a grid separation distance of 154.7~mm. Together with photon detectors, read-out electronics, and a data-processing unit, likely requiring an area 10-15~cm long, the Imager could fit into the length of 3U. 

If gratings of as small as 15 \micron{} can be made with the quality required to produce the needed \moire{} pattern, it should be less challenging to make slightly coarser gratings that a larger separation distance would allow.  
For example, a separation of 206.3~mm would require 120~\micron{}
 period for the 60~arcsec angular scale and 20~\micron{} for 10~arcsec.

 The complete parameters characterizing the EM grids are shown in Table~\ref{tab-EM-param}.

   \begin{table}[h]
  \centering
  \resizebox{\textwidth}{!}{%
  \begin{tabular}{cccccccccc}
      & & &  & & & Front Grid & &Rear Grid& \\
     \bf Segment & \bf Subcoll. & \bf Fourier & \bf Sampled  & \bf Phase & \bf Angular  &\bf Period (p) &\bf Angle ($\pm\alpha$)&\bf Period (q) &\bf Angle ($\pm\beta$) \\
     \bf  & \bf  & \bf N& \bf  dir. & \bf  & \bf  res. (arcsec)&\bf ($\mu$m) &\bf (deg) &\bf ($\mu$m) &\bf (deg.) \\
    
     \hline
     B & 11 & 1 & $-45^\circ$ & $+1$ & 60 & 89.675	&-44.794&	90.326	&-45.208 \\
     B & 21 & 1 & $+45^\circ$ & $+1$ & 60 &	90.326	&45.208&	89.675	&44.794\\
     B & 12 & 4 & $+45^\circ$ & $-1$ & 15 & 22.480	&44.948&	22.520	&45.052\\
     B & 22 & 4 & $-45^\circ$ & $-1$ & 15 & 22.520	&-45.0512&	22.480	&-44.948\\	
    &&&&&&&&\\
     C & 31 & 2 & $-45^\circ$ & $-1$ & 30 & 45.081	&-45.104&	44.919	&-44.897\\
     C & 41 & 2 & $+45^\circ$ & $-1$ & 30 & 44.919	&44.897&	45.081	&45.104\\
     C & 32 & 5 & $+45^\circ$ & $+1$ & 12 & 18.013	&45.0416&	17.987	&44.959\\
     C & 42 & 5 & $-45^\circ$ & $+1$ & 12 & 17.987	&-44.959&	18.013	&-45.041\\
    &&&&&&&&\\
     D & 33 & 3 & $-45^\circ$ & $+1$ & 20 & 29.964	&-44.931&	30.036	&-45.069\\
     D & 43 & 3 & $+45^\circ$ & $+1$ & 20 & 30.036	&45.0691&	29.964	&44.931\\
     D & 34 & 6 & $+45^\circ$ & $-1$ & 10 & 14.991	&44.965&	15.009	&45.035\\
     D & 44 & 6 & $-45^\circ$ & $-1$ & 10 & 15.009	&-45.035&	14.991	&-44.966\\
      \hline
  \end{tabular}}
  \caption{MiSolFA EM parameters, with separation distance 154.7 mm and Moir\'e period 8.8 mm.}
  \label{tab-EM-param}
 \end{table}

The minimum grating pitch, together with requirement that the grating be thick enough ($\approx >$200 \micron) to stop photons in the desired energy range, drove the choice of manufacturing technique, which in turn limited the material choice for the gratings.
Neither of the grid production techniques employed by RHESSI or STIX can make a grating of this thickness with periods less than 34 \micron, as in STIX's finest tungsten gratings. Their thickness of 120 \micron{} was achieved by stacking thin ($\approx$30 \micron) foils that had been previously etched. The smallest features that can be produced by chemical etching have similar size to the foil thickness and can depend on the grain structure of the metal layer. 

New etching-based techniques (metal-assisted chemical etching: \citet{romanoSelfassemblyNanostructuredGold2016,romanoEffectIsopropanolGold2017}, Bosch etching with seedless gold electroplating: \citet{jefimovsHighaspectRatioSilicon2017,kagiasFabricationAuGratings2019}) were attempted but failed to produce prototype gratings of functional thickness. However, success was found with the LIGA (X-ray lithography, electroplating and moulding) method employed by MicroWorks GmbH. LIGA boasts the capability to produce tall, straight microstructures with well-defined geometry and smooth sidewalls, all within very tight tolerances \citep{meyerChapter16Deep2015}. For MiSolFA, MicroWorks used a carbon substrate on which they manufactured gold slats with a thickness of more than 200~\micron, far outperforming etching methods.

\begin{figure}[t]
   \includegraphics[width=.49\textwidth]{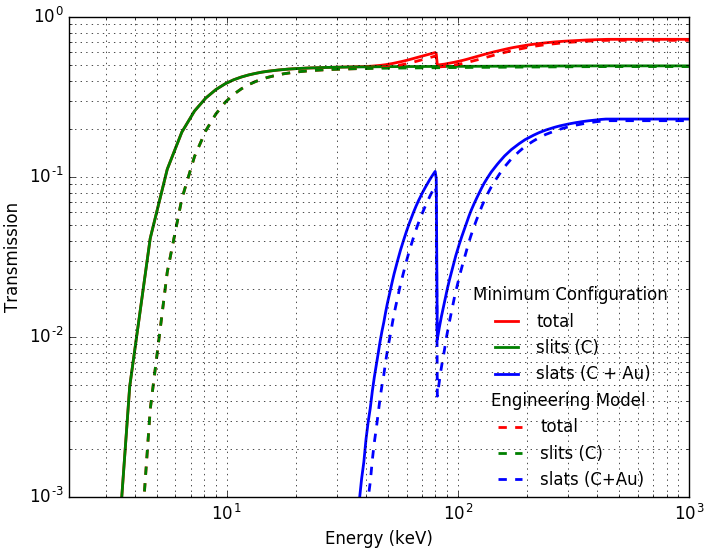}
   \includegraphics[width=.5\textwidth]{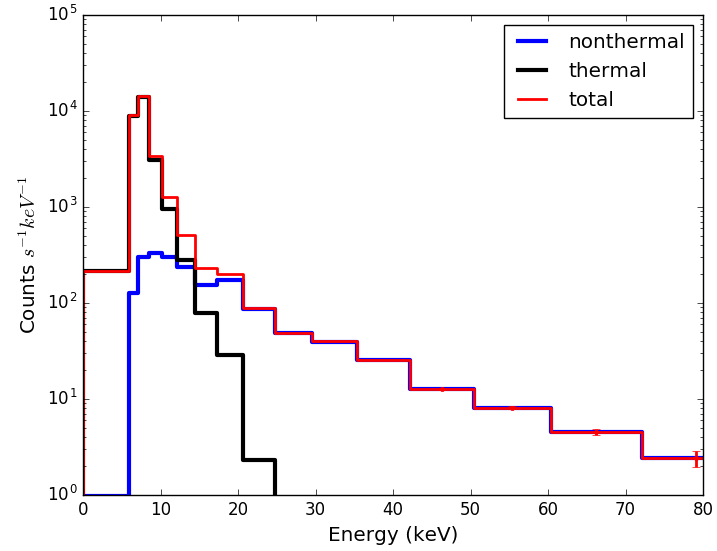}
   \caption{Left: Transmission (percent) for the materials used to construct a single grating. The desired thickness of the gold slats is 250 $\mu$m (minimum configuration thickness is 200 \micron) and the thickness of the carbon substrate is 1 mm (minimum 0.8 mm). Right: Expected number of counts for a C-class flare on a detector area of 12 cm$^{2}$, for such a grating, using logarithmic binning to increase nonthermal counts.}
   \label{fig-att-len}
 \end{figure}

Figure \ref{fig-att-len} (left side) shows the calculated transmission of the carbon-and-gold grating design. 
The nominal gold thickness is 250 $\mu$m, with the minimum configuration of 200 \micron{} thickness still providing close to the ideal 50\% transmission (half of the photons should be absorbed by the gold slats) between 10 and 100 keV.  
On the right, the expected number of counts through such a grating pair is shown given the input distribution of the C-class flare of 20 February, 2002 \citep{kruckerRelativeTimingSpectra2002a}. With the use of logarithmic binning, there is a sufficient number of counts to quantify the nature of the non-thermal component. While MiSolFA will certainly be able to detect soft X-ray emission from smaller flares, this is less relevant to the science goals. Extremely bright X-class flares might result in pileup or livetime effects, so a fixed  attenuator of absorbing material could be wise addition to the instrument to optimize the design for larger flares.    


\begin{figure}
\centering
\includegraphics[width=.6\textwidth]{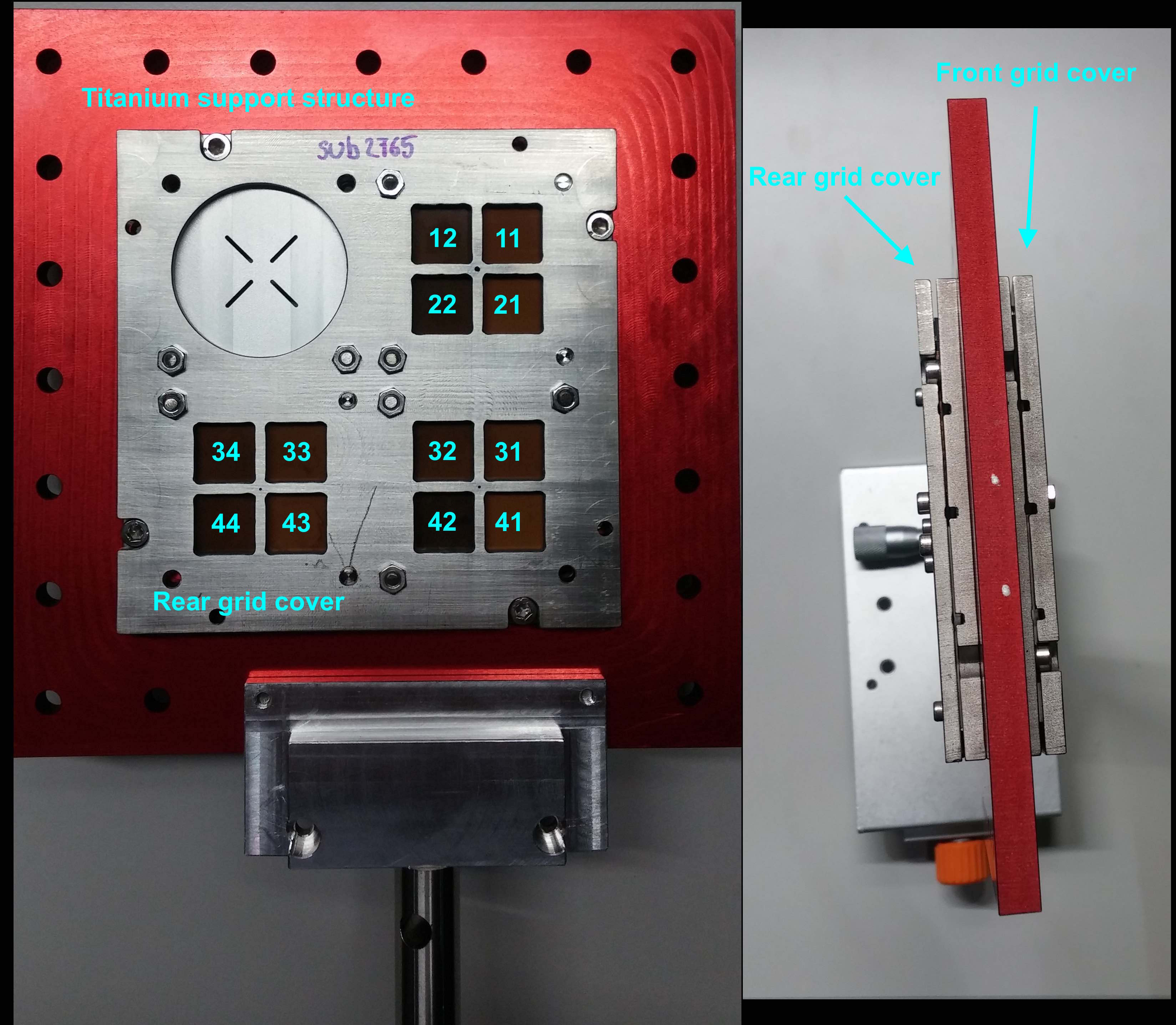}
\caption{Rear grid (left) and top (right) view of the assembled EM Imager. The segments containing the windows are placed in the titanium covers, which are then attached to a support. Each grid assembly can be removed from the support at will.}
\label{fig:EMimager}
\end{figure}

An Engineering Model of the MiSolFA Imager was assembled using prototype gratings in April 2017. Segment B was on a separate substrate from Segments C and D, so each grid (front and rear) was assembled by gluing the individual substrates into specially designed titanium covers. During this process, a crack developed in the front (Sun-side) grid's Segment C. The complete front and rear grid assemblies could either be treated separately or screwed back-to-back onto a support, forming a single unit as shown in Figure \ref{fig:EMimager}.

\section{Performance of the EM Imager}\label{sec-perf}

An operational Imager must, when exposed to X-ray emission from solar flares, produce \moire{} patterns of sufficient uniformity to be well-characterized by the detectors.  
Therefore, the performance of the EM Imager was evaluated based on the moiré patterns it produced. We required that together the front and rear grid produce detectable moiré patterns that remained consistent after the completion of space qualification tests. An EM that can meet these criteria gives us reasonable expectation that such gratings could operate in space in a way that meets the science goals of the MiSolFA mission.

\subsection{Space qualification tests}
To be considered space qualified, the grids must pass vibration and thermal cycling tests. We chose to use the undamaged rear grid for the qualification tests.

The objective of the vibration test was to demonstrate the structural integrity of the rear grid and show that the quality of the moiré pattern obtained by the combination of front and rear grid was unaffected after the test. To achieve this, a typical launch environment ($g_{RMS}$ from 11g to 16g) experienced by a scientific payload mounted on a satellite panel was simulated by an electrodynamic shaker generating sine and random vibration loads in both in-plane directions as well as in the out-of-plane direction.

The thermal test of the grid was designed to judge if the temperature environment experienced during launch and operation would affect its performance. The grid was placed in a thermal chamber and underwent temperature cycles starting at 20\degree C and ranging from -30\degree to +70\degree C, with the change in temperature accomplished over one hour and plateauing at either extreme for an additional hour. Eight cycles were performed, with additional time at the beginning and end to stabilize to 20\degree C. Ice crystals forming from the moisture in the air, upon expansion, posed the greatest threat of damage the microstructures of the gratings. Because this was not enough to damage the grid, we expect that a similar test in a vacuum chamber would be passed successfully.

Following each of these tests, no visible structural damage was observed, and no change in phase (indicative of damage) was noted in the post-test moiré patterns. \emph{We conclude, in short, that the space environment will not affect the performance of the MiSolFA EM Imager.}

\subsection{Performance Assessment via Moir\'e patterns}

We measured the \moire{} patterns produced by the EM multiple times: first, after the EM gratings were first received, and then once after each space qualification test. Ideally, this would be done using a plane-parallel synchrotron X-ray source to better mimic the Sun, but due to cost and scheduling issues we used a tabletop X-ray lamp, which produces divergent X-rays. In this section, we discuss the set of measurements taken after the vibration test and thermal test were completed. This is the most important set of images, as effects of any stress from the space qualification tests should be seen.

The experimental setup had the front and rear grids placed back-to-back as shown in Figure \ref{fig:EMimager}. The assembly was positioned about a meter from the source, oriented so that the subcollimators of one segment would form patterns within the field of view of the detector plate. With the setup employed, this usually resulted in the assembly being located less than 10 cm from the detector. If needed, the assembly was rotated or translated such that moiré patterns of the desired segment appeared on the detector. Since it was not possible to measure any of these distances or the rotation angle to a high degree of accuracy; in any case, the figure of merit in this setup is the uniformity of the \moire{} pattern which is not affected by these measurements.

\begin{figure}
    \centering
    \includegraphics[width=0.32\textwidth]{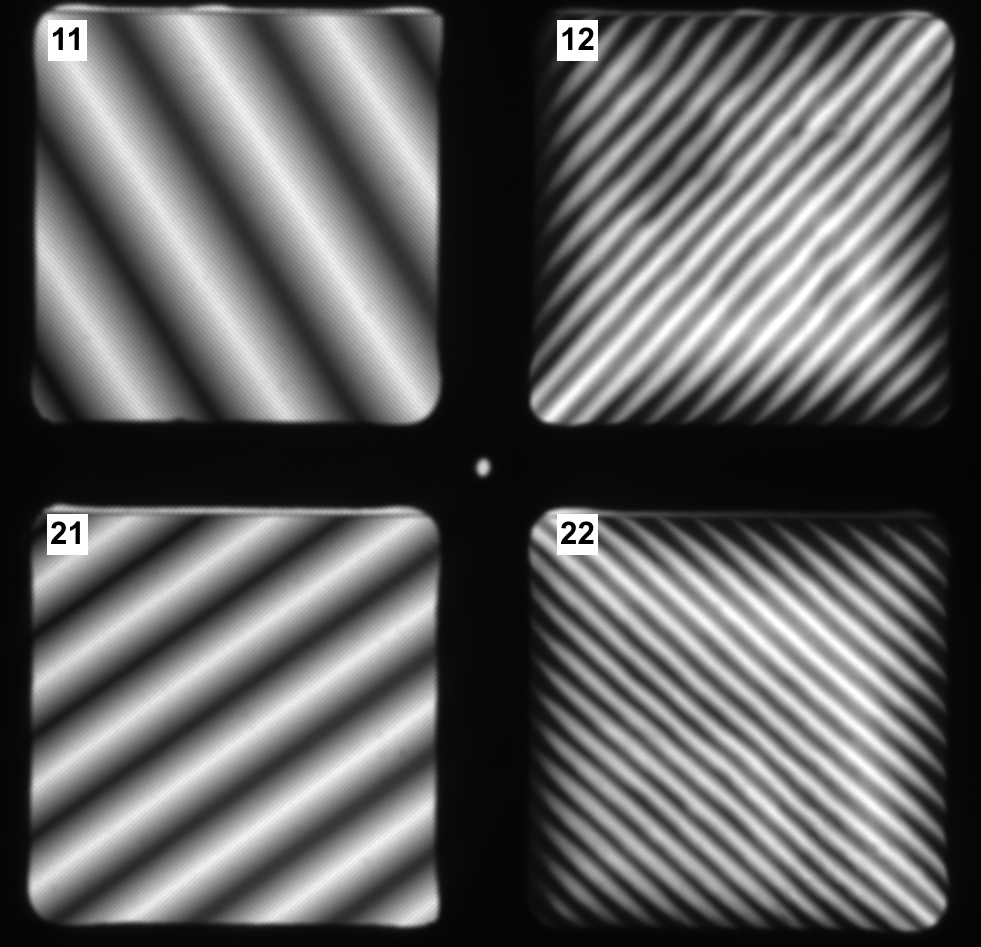}
    \includegraphics[width=0.32\textwidth]{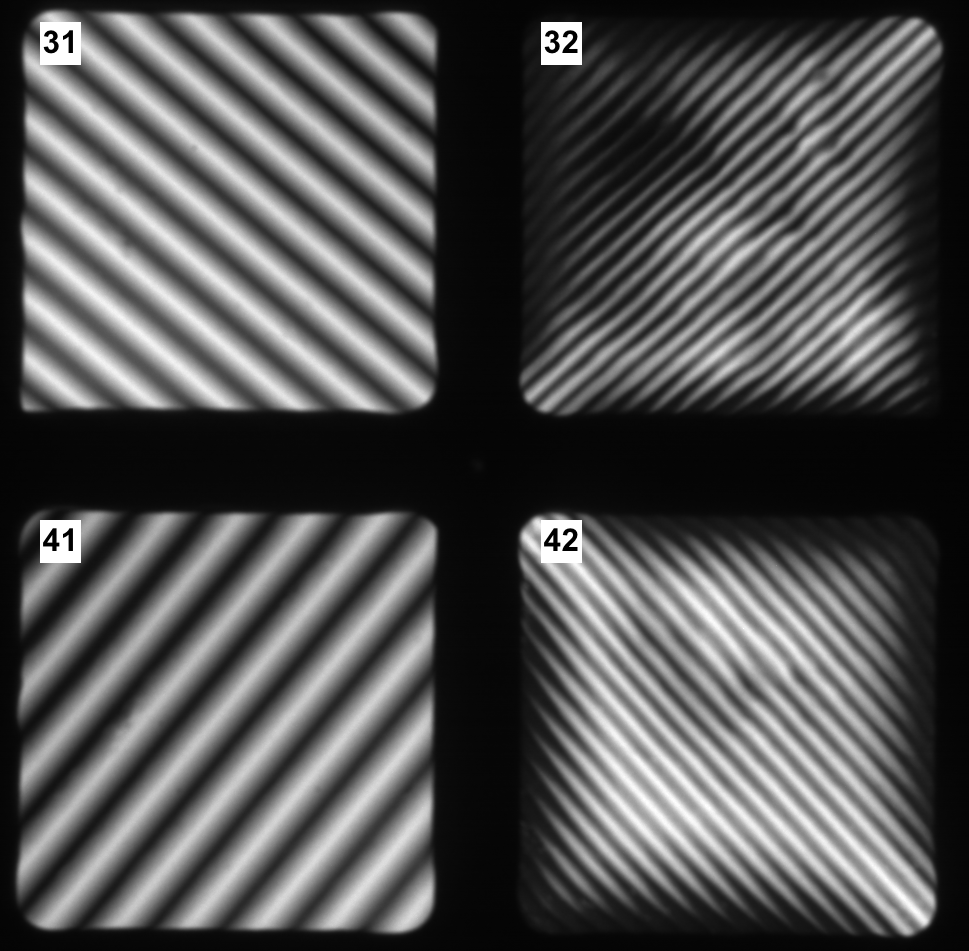}
    \includegraphics[width=0.32\textwidth]{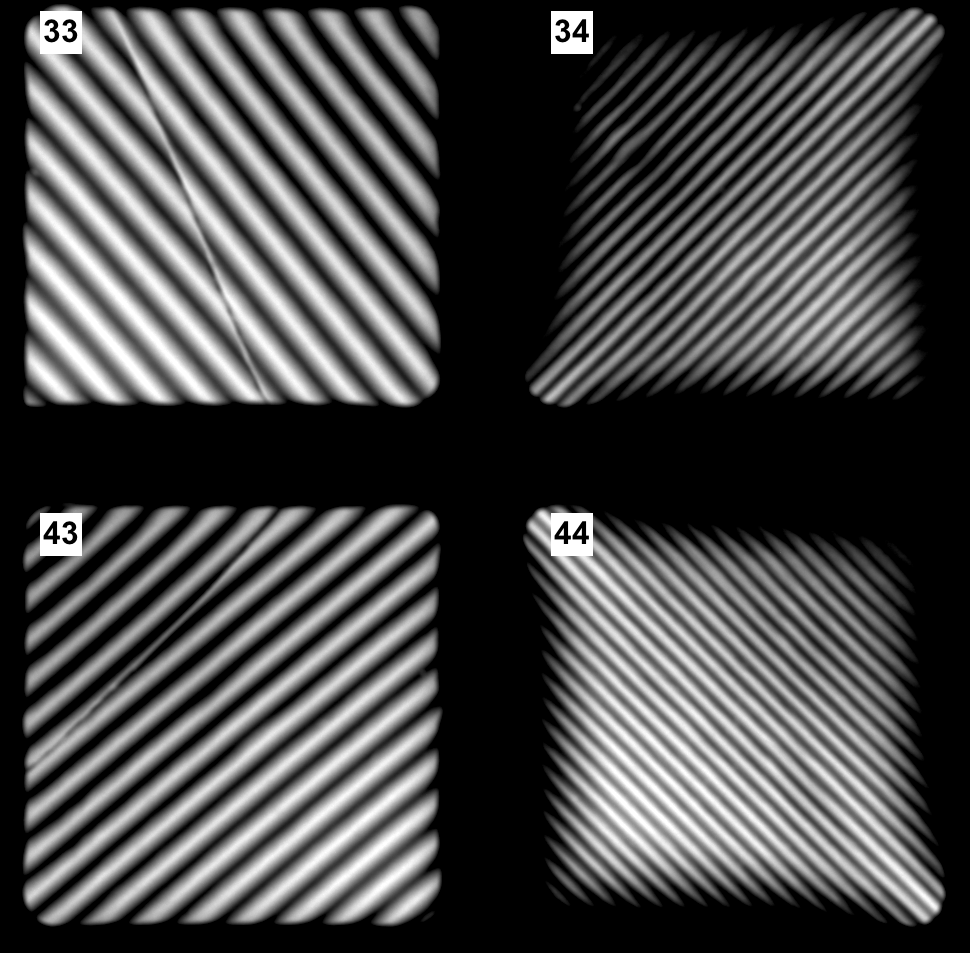}
    \includegraphics[width=0.33\textwidth]{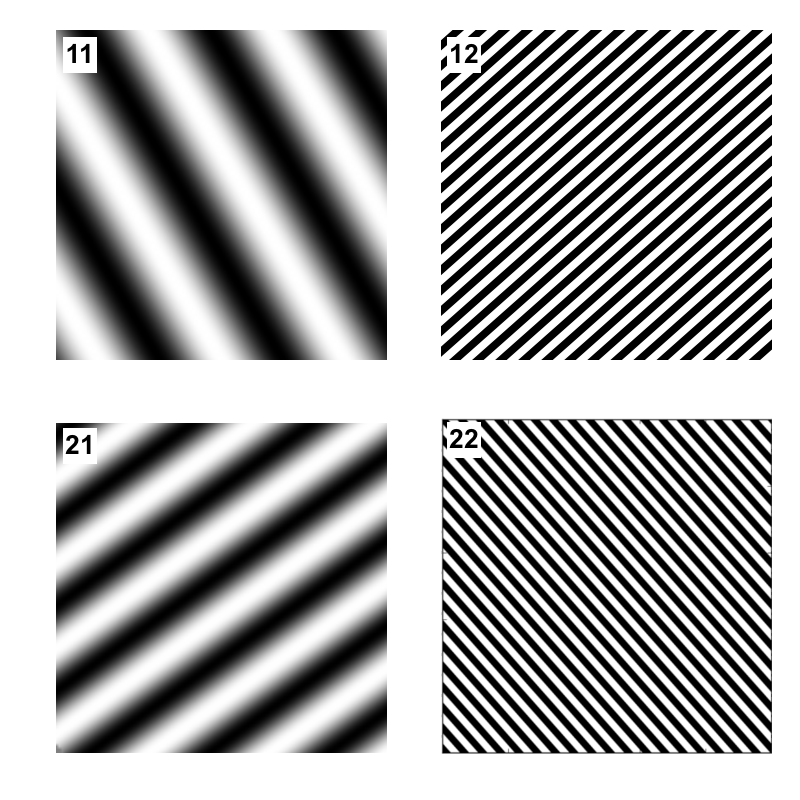}
    \includegraphics[width=0.66\textwidth]{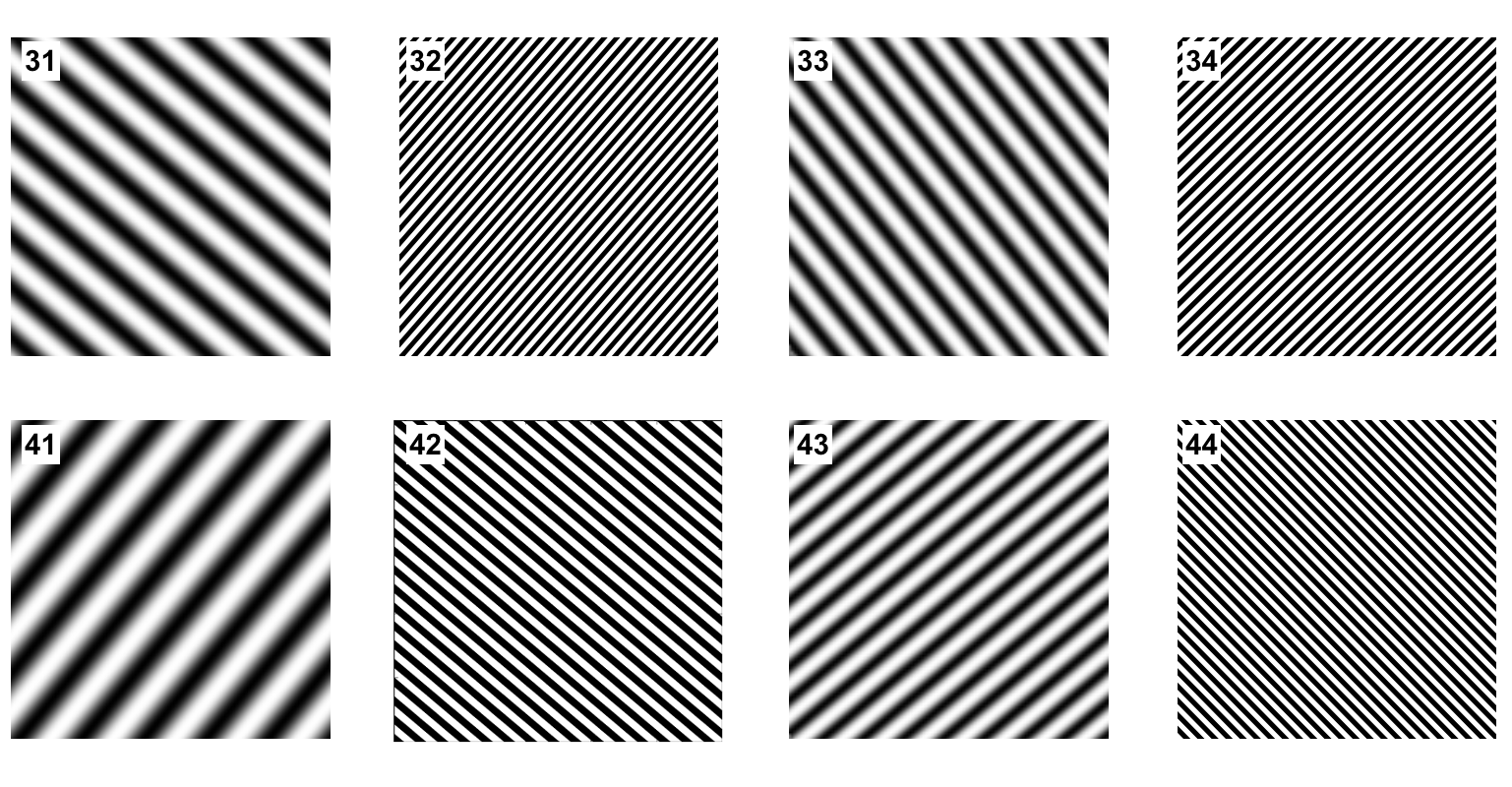}
    \caption{Top: Observed \moire{} patterns from an X-ray lamp test on 2 Feb 2018. Bottom: Patterns re-constructed from a least-squares fit to find the exact experiment parameters, assuming nominal grating values.}
    \label{fig:lamptest}
\end{figure}

The \moire{} patterns observed (Figure \ref{fig:lamptest}, top row) display great overall uniformity, with local defects prominent in only the finer-pitched subcollimators. The darker patches affecting the patterns indicated that the transmission of the gratings varies from place to place. An initial examination with an optical microscope, performed before the space qualification tests, revealed several potential causes for such variations. In a few places, parts of the gold slats were missing, and in others a small groups appeared pinched together. The borders of the gratings were particularly subject to variation in the lengths and local orientations of the slats. This, combined with shadowing due to the divergent X-ray source used in obtaining the \moire{} images, accounts for the edge-darkening observed in the finer-pitched windows. Local changes to grating period and orientation result in a local phase shift of the \moire{} pattern, as dramatically demonstrated by the cracked windows, 33 and 43. On a smaller scale, this means an intensity change in areas that become out-of-phase with the greater pattern. Defects like these, once well characterized, can be corrected for during data analysis. 

Large changes in the period and orientation of the \moire{} patterns depend mainly on the geometry of the experimental setup and on one additional factor - the unknown relative twist between the front and rear grid segments. We did not know if segment B in the front grid was rotated with respect to segment B in the rear grid, due the position of the segment inside the grid cover or the attachment of the covers and gratings to the support structure. This can very easily change the \moire{} pattern, as it effectively changes the orientation of the gratings in each segment by some amount. It is important to distinguish this twist, which is different for Segment B and Segments C and D, from actual variation in the orientations of individual gratings. Previous studies with STIX and RHESSI have shown that it is possible to achieve good twist alignment in the instrument itself.

Once the lamp images were obtained, we measured the moiré period and orientation by rotating the X-ray image of each window until the bright and dark stripes of the pattern were as vertical as possible, determined by the rotation angle that maximized the gradient of the sum. This rotation angle was taken to represent the \moire{} pattern orientation. Period was calculated by finding the distance between the relative maxima. For windows 33 and 43, affected by the crack, we used the patterns seen on the innermost halves of the windows for the characterization.

Next, we used the measured \moire{} pattern properties together with the nominal grating parameters in Table \ref{tab-EM-param} to determine if the observed patterns were indeed close to what we would expect from perfect gratings. This calculation is easier if one can assume plane-parallel X-rays, but even with a divergent X-ray source like the R\"ontgen lamp, it can be done using the geometric approximation. This small-angle assumption is safe to make because the distance between the source and the imager is orders of magnitude larger than a single slit width in even the coarsest grating. 
We also assumed the light was monochromatic and originated from a point source, and treated the gratings themselves as having zero thickness, ignoring scattering and diffraction.

When illuminated with divergent X-rays, a slit in the front grating with width $w$  becomes projected onto the second grating as a magnification $w_p = ws$, where the geometric characteristic of the layout is $s = d_2/d_1$ or $s=1+(d_2 - d_1)/d_1$. Here, $d_1$ is the distance between the source and the front grid, and $d_2$ is the distance between the front and the rear grid. 
Assuming that the detector plane is at also the position of the second grid, we can easily calculate the resulting moire pattern period and orientation in the usual fashion, using the projected period $w_p$ as the period of the first grating. 
To find the parameters which best reconciled the observed \moire{} patterns with the experimental setup, we then performed a least-squares fit to find $d_1$, $d_2$, and the relative twist, restricting the distances to within a centimeter of what was measured and the relative twist to less than one degree. Because of the way the segments were manufactured, we fit segment B, which was made on one substrate, separately from segments C and D.

The observed patterns and those calculated from the best-fit parameters are shown in Figure \ref{fig:lamptest}. Relative twist was almost negligible, between $0.001 - 0.008\degree$ for segment B and $0.056\degree$ for segments C and D. 
We found that with only very small adjustments, we could easily reproduce the \moire{} patterns observed experimentally. This indicates that the gratings as a whole do not deviate a significant amount from the nominal parameters requested of the manufacturers. 
\emph{In conclusion, the gratings produced for the MiSolFA EM are capable of producing \moire{} patterns that are both regular enough and close enough to nominal to perform in a way that will accomplish MiSolFA's science goals.}

\section{Imager Current Status}

Although the X-ray lamp images provided sufficient evidence that the EM gratings would perform well, we also collected a comprehensive set of optical and synchrotron X-ray images of each individual grating. Using this data, we can obtain the RMS deviations from the nominal period and orientation of each grating, allowing us ultimately to derive the grating efficiency \citep{hurfordRHESSIImagingConcept2002}. These numbers, if well known, are used during image reconstruction to mitigate imperfections in the gratings. Previous work has shown that even seemingly large but localized deviations in the period, which lead to effectively deviations in the phase, will only lower the efficiency by a small percentage. Thanks to the segmented nature of the grids, we have the flexibility to mitigate a bad RMS in the orientation by tuning the relative twist of the front and rear gratings to minimize the effects of any offsets in the ‘worst’ grating in each segment. Based on this knowledge and initial inspections, we are confident that the EM gratings will be reasonably efficient – certainly enough to reconstruct science-worthy images once all possible defects are accounted for. Full characterization is  currently being performed but is out of scope of this paper.

A second Engineering Model was produced in 2018. Some small design changes were implemented that changed the nominal \moire{} patterns slightly; however, the most noticeable change was that the gratings were all manufactured on a single, monolithic substrate. In theory, this simplifies alignment by reducing the twist parameter to as single angle; in practice, the electroplating caused internal stresses on the graphite substrate that resulted in both a noticeable convexity towards the center of the piece and an unfortunate radial crack in the front grid that developed during shipping. X-ray lamp images show that the fine-pitched gratings produce \moire{} patterns that are less affected by defects than the gratings discussed in the previous section, indicating a marked improvement in their quality. Like with the first set of EM gratings, optical and X-ray characterization of the transmission and efficiency is ongoing.

\section{Conclusion}

We have shown that the MiSolFA concept is a viable way to miniaturize the key components of an indirect X-ray imager. 
Advances in grating production technology allowed us to realize the production Engineering Model grids that are good enough to operate in space.
By working together, STIX and MiSolFA can produce one-of-a-kind stereoscopic high-energy observations, proving that even a micro-satellite a can contribute to solving fundamental problems in solar physics.

\section*{References}
\bibliographystyle{agsm}
\bibliography{directivity}

 \end{document}